%% file: 7technology.tex
\documentclass[11pt,twocolumn]{article}
\usepackage{amssymb,epsfig,wrapfig,graphics,aas_macros}
\usepackage[dvips]{color}
\usepackage[footnotesize]{caption}

\def\comment#1{{}}

\newlength{\cvindent}
\newlength{\cvhang}

\newlength{\refindent}
\begin{document}

\vspace{-0.5in}

\title{The Status and future of ground-based TeV gamma-ray astronomy\\
Reports of Individual Working Groups}
\date{}
\maketitle
\section{Technology}
\label{grb-subsec}
\input{7-technology.tex}

\end{document}

%% file: 7-technology.tex
Group membership:\\ \\
K. Byrum, J. Buckley, S. Bugayov, B. Dingus, S. Fegan, S. Funk, E. Hays, J. Holder, D. Horan, A. Konopelko,  H. Krawczynski, F. Krennrich, S. Lebohec, G. Sinnis, A. Smith, V. Vassiliev, S. Wakely

\subsection[Introduction and overview]{Introduction and Overview}

High-energy gamma rays can be observed from the ground by detecting 
secondary particles of the atmospheric cascades initiated by the interaction 
of the gamma-ray with the atmosphere. Imaging atmospheric Cherenkov 
telescopes (IACTs) detect broadband spectrum Cherenkov photons ($\lambda > 
300$ nm), which are produced by electrons and positrons of the cascade and 
reach the ground level without significant attenuation.    
The technique utilizes large mirrors to focus Cherenkov photons onto a finely 
pixelated camera operating with an exposure of a few nanoseconds, and 
provides low energy threshold and excellent calorimetric capabilities. The 
IACTs can only operate during clear moonless and, more recently, partially-moonlit nights. Alternatively, the 
extended air shower (EAS) arrays, which directly detect particles of the 
atmospheric cascade (electrons, photons, muons, etc.) can be operated 
continuously but require considerably larger energy of the gamma rays 
necessary for extensive air showers to reach the ground level.

The field of TeV gamma-ray astronomy was born in the years 1986 to 1988 
with the first indisputable detection of a cosmic source of TeV gamma
rays with the Whipple $10$~m IACT, the Crab Nebula 
\cite{1989ApJ...342..379W}. Modern IACT observatories such as VERITAS 
\cite{Week:02,Maie:07}, MAGIC \cite{2004NewAR..48..339L,Goeb:07}, and 
H.E.S.S. \cite{2004NewAR..48..331H,Horn:07} can detect point sources with a 
flux sensitivity of $1\%$ of the Crab Nebula corresponding to a limiting $\nu 
$F$_{\nu }$-flux of $\sim 5\times 10^{-13}$ ergs cm$^{-2}$ s$^{-1}$ at 1 TeV. 
The improvement of sensitivity by two orders of magnitude during the last two 
decades has been made possible due to critical advances in IACT technology 
and significantly increased funding for ground-based gamma-ray astronomy. 
The high point-source flux sensitivity of IACT observatories is a result of  
their large gamma-ray collecting area ($\sim 10^{5}$ m$^{2}$), relatively 
high angular resolution ($\sim 5$ arcminutes), wide energy coverage (from 
$<100$ GeV to $>10$ TeV), and unique means to reject cosmic ray background 
($> 99.999\%$ at 1 TeV). The limitations of the IACT technique are the small 
duty cycle ($\sim 10\%$), and narrow field of view ($\sim 4\deg $; $3.8\times 
10^{-3}$ sr for present-day IACTs).

Large EAS arrays provide complementary technology for observations of very 
high-energy gamma rays. Whereas their instantaneous sensitivity is 
currently a factor $\sim 150$ less sensitive than that of IACT observatories, their 
large field of view ($\sim 90\deg $; $1.8$ sr) and nearly $100\%$ duty cycle 
makes these observatories particularly suited to conduct all-sky surveys and 
detect emission from extended astrophysical sources (larger than 
$\sim 1\deg $, e.g. plane of the Galaxy). Milagro \cite{Smit:05}, the first 
ground-based gamma-ray observatory which utilized EAS technology to 
discover extended sources \cite{Abdo:07}, has surveyed $2\pi $~sr of 
the sky at $20$~TeV for point sources to a sensitivity of $3\times 10^{-12}$ 
ergs cm$^{-2}$ s$^{-1}$. Due to the wide field of view 
coverage of the sky and uninterrupted operation, the EAS technique also has 
the potential for detection of Very High Energy (VHE) transient phenomena. 
The current limitations of EAS technique are high-energy threshold ($\sim 10$ 
TeV), low angular resolution ($\sim 30$ arcminutes), and limited capability to reject cosmic-ray background and measure energy.

The primary technical goal for the construction of the next generation of observatories is to 
achieve an improvement of sensitivity by a factor of $\alpha $ at the cost 
increase less than a factor of $\alpha ^{2}$, the increase that would be required if the observatory were constructed by simply cloning present day instrumentation~\footnote{Background dominated regime of observatory 
operation is assumed}. The history of ground-based gamma-ray astronomy 
over the last two decades has shown twice an improvement in the sensitivity of the observatories by a factor of ten while the cost has increased each time only by a factor of ten  
\cite{2007ebhe.conf..282W}. 

The construction of a large array of IACTs covering an area of $\sim 1$ km$^2$ will enable ground-based $\gamma$-ray astronomy to achieve another order of magnitude improvement in sensitivity. This next step will be facilitated by several technology improvements. First, large arrays of IACTs should have the capability to operate over a broad energy range with significantly improved 
angular resolution and background rejection as compared to the present day 
small arrays of telescopes, such as VERITAS or H.E.S.S.. Second, the capability of using subarrays to fine tune the energy range to smaller intervals will allow for considerable reduction of aperture of 
individual telescopes and overall cost of the array while maintaining the 
collecting area at lower energies equal to the smaller array of very large 
aperture IACTs. Finally, the cost per telescope can be significantly reduced 
due to the advancements in technology, particularly the development of low cost 
electronics, novel telescope optics designs, replication methods for 
fabrication of mirrors, and high efficiency photo-detectors, and due to the 
distribution of initial significant non-recurring costs over a larger number of 
telescopes. 

In the case of EAS arrays, the breakthrough characterized by the 
improvement of sensitivity faster than the inverse square root of the array 
footprint area is possible due to mainly two factors. First, next generation 
EAS array must be constructed at a high elevation ($>4000$ m) to increase the number of particles in a shower by being closer to the altitude where the shower has the maximum number of particles. Thus, a lower energy threshold is possible and energy resolution is improved.  Second, the size of the EAS array needs to be increased in order to more fully contain the lateral distribution of the EAS.  A larger array improves the angular resolution of the gamma-ray showers and also dramatically improves the cosmic ray background rejections.  The lateral distribution of muons in a cosmic ray shower is very broad, and identification of a muon outside the shower core is key to rejecting the cosmic ray background.

The science motivations for the next generation ground-based gamma-ray 
observatories are outlined in this document.  There are clear cost, reliability, maintenance, engineering, 
and management challenges associated with construction and operation of a future 
ground-based astronomical facility of the order $\sim $100M dollar scale. 
Detailed technical implementation of a future observatory will benefit from current and future R\&D efforts that will provide better understanding 
of the uncertainties in evaluation of the cost impact of 
improved and novel photon detector technologies and from the current incomplete  
simulation design studies of the large optimization space of parameters of 
the observatory. 
In the remainder of this section, we outline a broadly defined technical roadmap for 
the design and construction of future instrumentation which could be realized within the next decade.  We start with a status of the field, 
identify the key future observatory design decisions, technical drivers, 
describe the current state of the art technologies, and finally outline a plan for 
defining the full technology approach.

\subsection{Status of ground-based gamma-ray observatories}{Status of Ground-Based Gamma-ray\\Observatories}
\begin{figure*}[t]
\begin{center}

\includegraphics[angle=0,width=6.0in]{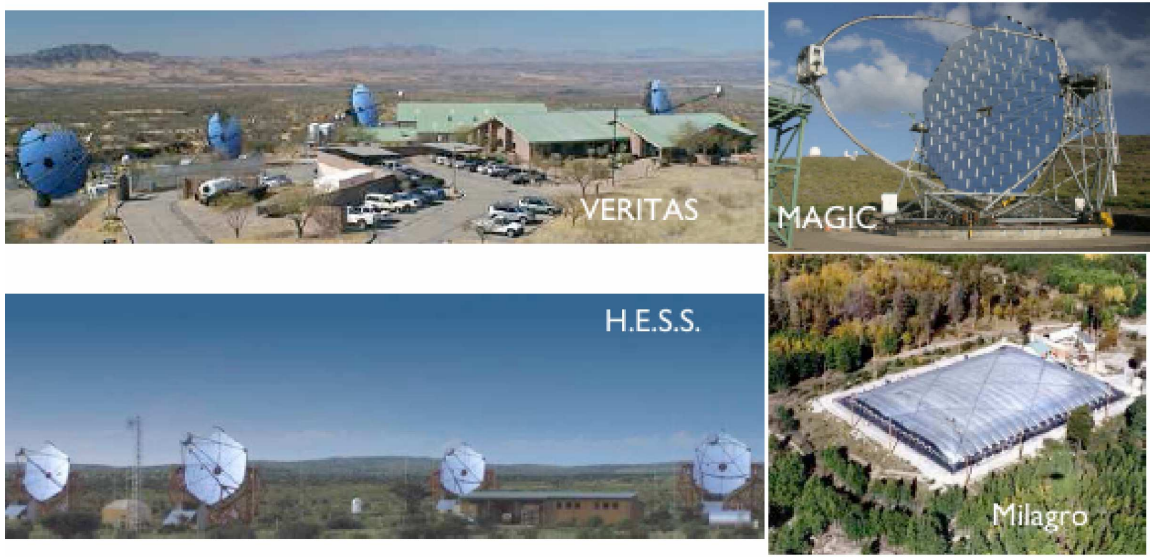}

\caption{\label{fig:exp} The images show four major ground-based 
gamma-ray observatories currently in operation: VERITAS, MAGIC, H.E.S.S.\ 
, and MILAGRO. A future ground-based gamma-ray project can build on the 
success of these instruments.}
\end{center}

\end{figure*}

At present, there are four major IACT and three EAS observatories worldwide conducting routine astronomical observations, four of which are shown in Fig \ \ref{fig:exp}.  Main parameters of these instruments are the following:

\paragraph{VERITAS} is a four-telescope array of IACTs located at the Fred Lawrence Whipple Observatory in Southern Arizona (1268 m a.s.l.). Each telescope is a 12 m diameter Davies-Cotton (DC) reflector (f/1.0) and a high resolution 3.5$\deg$ field of view camera assembled from 499 individual photo multiplier tubes (PMTs) with an angular size of 0.15 deg.  The telescope spacing varies from 35~m to 109~m.  VERITAS was commissioned to scientific operation in April 2007.

\paragraph{The H.E.S.S.\  array} consists of four 13 m DC IACTs (f/1.2) in the Khomas Highlands of Namibia (1800 m a.s.l.). The 5 deg field of view cameras of the telescopes contain 960 PMTs, each subtending 0.16deg angle.  The current telescopes are arranged on the corners of a square with 120m sides. H.E.S.S.\  has been operational since December 2003. The collaboration is currently in the process of upgrading the experiment (H.E.S.S.\ -II) by adding a central large (28 m diameter) telescope to the array to lower the trigger threshold for a subset of the events to 20 GeV and will also improve the sensitivity of the array above 100 GeV. 

\paragraph{MAGIC} is a single 17 m diameter parabolic reflector (f/1.0) located in the Canary Island La Palma (2200 m a.s.l.). It has been in operation since the end of 2003. The 3.5 deg non-homogenous camera of the telescope is made of 576 PMTs of two angular sizes 0.1deg (396 pixels) and 0.2deg (180 pixels). The MAGIC observatory is currently being upgraded to MAGIC-II with a second 17-m reflector being constructed 85 m from the first telescope. The addition of this second telescope will improve background rejection and increase energy resolution. 

\paragraph{CANGAROO-III} consists of an array of four 10 m IACTs (f/0.8) located in Woomera, South Australia (160 m a.s.l.) \cite{Mori:07}.  The telescope camera is equipped with an array of 552 PMTs subtending an angle of 0.2deg each.  The telescopes are arranged on the corners of a diamond with sides of 100 m.

\paragraph{Milagro} is an EAS water Cherenkov detector located near Los Alamos, New Mexico (2650 m a.s.l.). Milagro consists of a central pond detector with an area of 60 x 80m$^2$ at the surface and has sloping sides that lead to a 30 x 50 m$^2$ bottom at a depth of 8 m. It is filled with 5 million gallons of purified water and is covered by a light-tight high-density polypropylene line.  Milagro consists of two layers of upward pointing 8'' PMTs.  The tank is surrounded with an array of water tanks. The central pond detector has been operational since 2000. The array of water tanks was completed in 2004.

\paragraph{The AS-\large{$\gamma$} and ARGO arrays} are located at the YangBaJing high-altitude laboratory in Tibet, China.  AS-$\gamma$, an array of plastic scintillator detectors, has been operational since the mid 1990s. ARGO consists of a large continuous array of Resistive Plate Counters (RPCs) and will become operational in 2007 \cite{Zao:05}.

\bigskip

The current generation of ground based instruments has been joined in mid-2008 by the space-borne \textbf{Fermi Gamma-ray Space Telescope} (formerly GLAST). Fermi comprises two instruments, the Large Area Telescope (LAT) \cite{McEn:07} and the Fermi Gamma-ray Burst Monitor (GBM) \cite{Lich:07}. The LAT covers the gamma-ray 
energy band of 20 MeV - 300 GeV with some spectral overlap with IACTs.  The present generation of IACTs 
match the $\nu F_{\nu}$-sensitivity of Fermi. Next-generation ground-based observatories with one order of 
magnitude higher sensitivity and significantly improved angular resolution would be ideally suited to 
conduct detailed studies of the Fermi sources.

\begin{table*}[!ht]
\begin{center}
\caption{\label{regimes} Gamma-ray energy regimes, scientific highlights and 
technical challenges.}

{\footnotesize

\begin{tabular}{p{0.6in} p{0.6in} p{1.9in} p{2.6in}}

\hline\hline

Regime & Energy Range & Primary Science Drivers & Requirements/Limitations \\ 
\hline

{\bf multi-GeV}: & $\leq$50~GeV &
extragalactic sources (AGN, GRBs) at cosmological distances ($z>1$),
Microquasars, Pulsars  &
very large aperture or dense arrays of IACTs, preferably high altitude operation \& 
high quantum efficiency detectors
required; 
 angular resolution and energy resolution will be limited by shower 
fluctuations, cosmic-ray background rejection utilizing currently available technologies is inefficient.

\\

{\bf sub-TeV}: & 50~GeV -- 200~GeV & extragalactic sources at intermediate redshifts($z < 1$), search for 
dark matter, Galaxy Clusters, Pair Halos, 
Fermi sources &  very-large-aperture telescopes or dense arrays of mid-size telescopes and high light detection efficiency required;
limited but improving with energy cosmic-ray background rejection based on imaging analysis. For gamma-ray bursts, high altitude EAS array.

\\ 

{\bf TeV}: & 200~GeV -- 10~TeV &
nearby galaxies (dwarf, starburst), nearby AGN, detailed 
morphology of extended galactic sources (SNRs, GMCs, PWNe) 
& large arrays of IACTs: best energy flux sensitivity, best angular and energy resolutions, best cosmic-ray hadron background rejection, new backgrounds from cosmic-ray electrons may ultimately limit sensitivity in some regions of the energy interval.  At the highest energy end, an irreducible background may be due to single-pion sub-showers. EAS arrays for mapping Galactic diffuse emission, AGN flares, and sensitivity to extended sources.

\\

{\bf sub-PeV}:  & $\geq$10~TeV & Cosmic Ray PeVatrons (SNRs, PWNe, GC, ...), 
origin
of galactic cosmic rays & 
requires very large (10 km$^2$ scale) detection areas; large arrays of IACTs equipped with very wide ($\ge 6^\circ$) FoV cameras and separated with distance of several hundred meters may provide adequate technology.  Background rejection is excellent and sensitivity is $\gamma$-ray count limited.  Single-pion sub-showers is ultimate background limiting sensitivity for very deep observations.  Regime of best performance of present EAS arrays; large EAS arrays ($\ge 10^{5}m^{2}$).

\\

\hline

\end{tabular}
}

\end{center}

\end{table*}

\subsection[Design considerations for a next-generation gamma-ray detector]{Design Considerations for a Next-Generation Gamma-Ray Detector}

At the core of the design of a large scale ground-based gamma-ray
 observatory is the requirement to improve the integral flux sensitivity by an order of magnitude over instruments employed today in 
the 50 GeV-20~TeV regime where the techniques are proven to give excellent 
performance. At lower energies (below 50 GeV) and at much higher energies 
(50-200 TeV) there is great discovery potential, but new technical approaches 
must be explored and the scientific benefit is in some cases less certain. 
For particle-detector (EAS) arrays, it is possible to simultaneously improve 
energy threshold and effective area by increasing the elevation, and the 
technical road-map is relatively well-defined.  In considering the design 
of future IACT arrays, the development path allows for complementary branches to more fully maximize the greatest sensitivity for a broad energy 
range from 10~GeV up to 100~TeV. 
Table \ref{regimes} summarizes specific issues of the detection technique and 
scientific objectives for four broad energy regimes (adapted from 
\cite{AharT:05,AharT:08}).
\subsection[Future IACT arrays]{Future IACT Arrays}

\begin{figure*}[t]

\begin{center}

\includegraphics[angle=0,width=3.in]{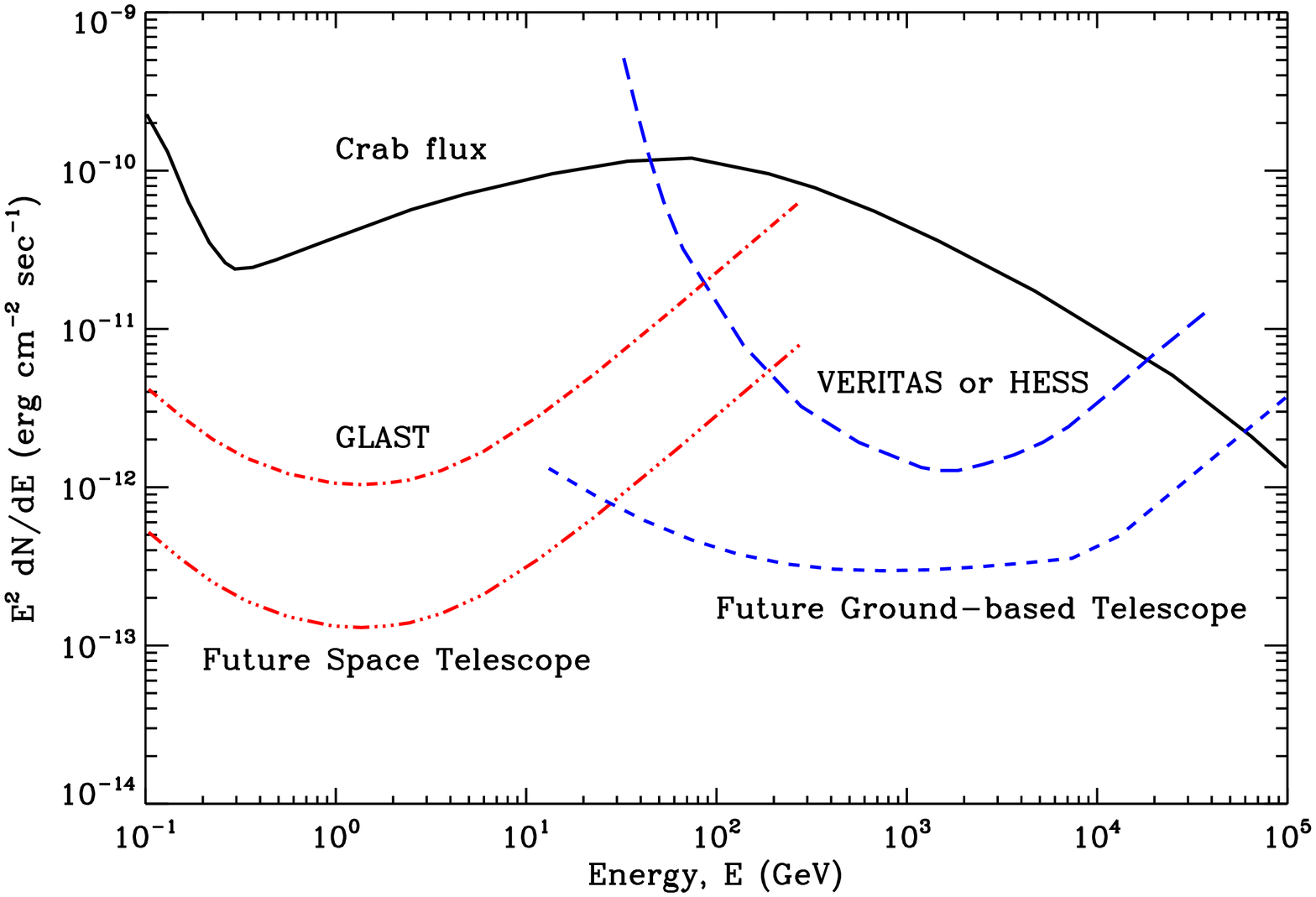}
\includegraphics[angle=0,width=3.in]{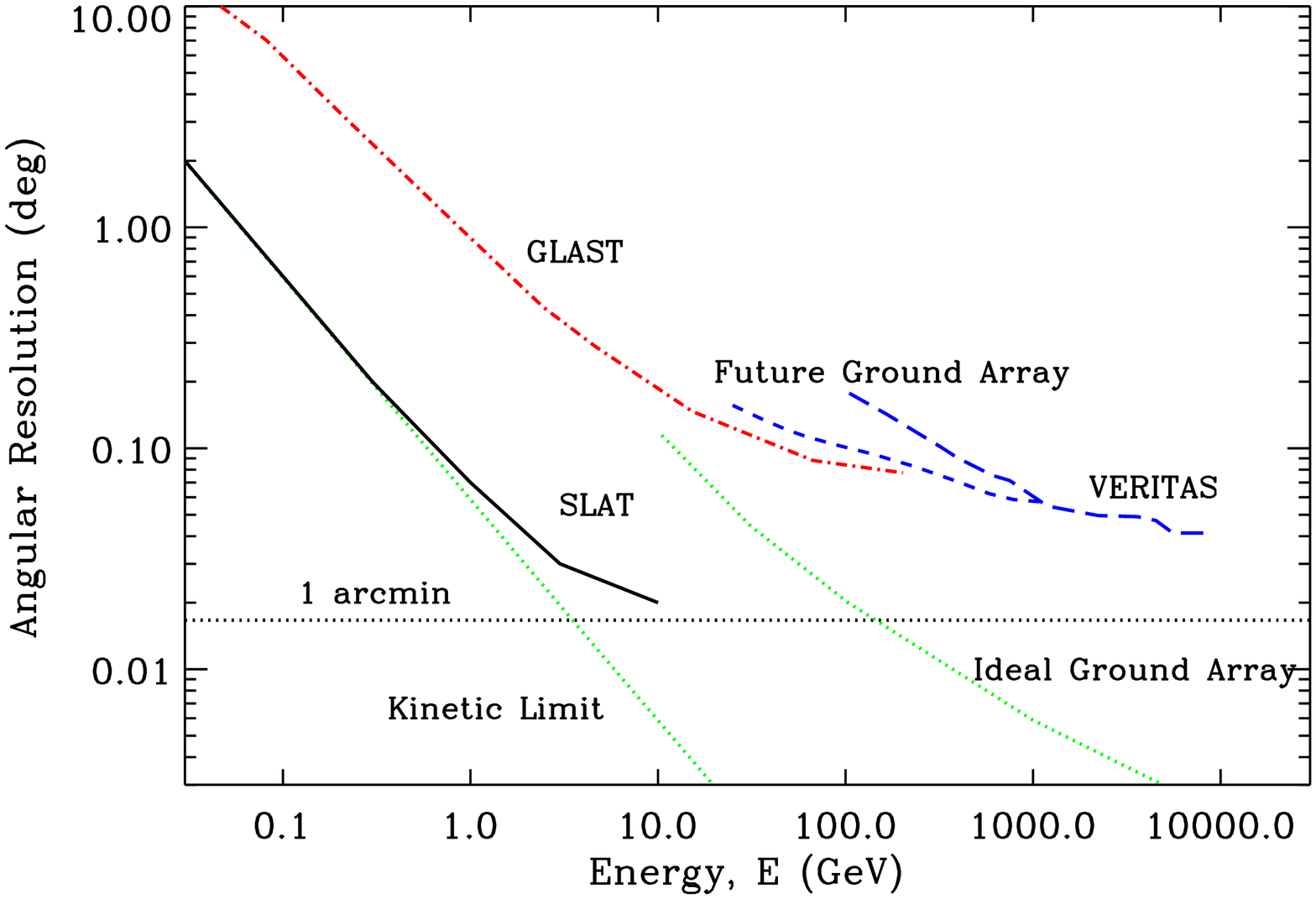}

\caption{ \textit{Left:} Differential sensitivities calculated for present and future gamma-ray experiments. For the future IACT array, an area of $\sim$1~km$^2$, no night-sky-background, a perfect point spread function \cite{Bugaev:07}, and an order of magnitude improvement in cosmic-ray rejection compared with current instruments has been assumed.  All sensitivities are 5 sigma detections in quarter decade energy intervals (chosen to be larger than the expected full-width energy resolution). \textit{Right} Angular resolution for Fermi (GLAST)
\cite{GLAST}, VERITAS \cite{Krawcz:06} and for ideal future space-borne and ground based \cite{Hofmann2005} gamma-ray detectors.}

\label{fig:sensang}

\end{center}

\end{figure*}

The scientific goals to be addressed with a future IACT array require 
a flux sensitivity at least a factor of ten better than  present-day observatories, 
and an operational energy range which extends preferably into the sub-100 GeV domain 
in order to open up the $\gamma$-ray horizon to observations of cosmologically distant 
sources. These requirements can be achieved by an array with a collecting area of 
$\sim 1$~km$^2$ (see Fig 1).  

The intrinsic properties of a $\sim 1$~km$^2$ IACT array could 
bring a major breakthrough for VHE gamma-ray astronomy since it combines several key
 advantages over existing 4-telescope arrays: 

\begin{itemize}

\item A  collection area that is 20 times larger than that of
existing arrays. Comparison of the collection area of a  $\sim 1$ km$^2$ array 
with the characteristic  size of the Cherenkov light pool ($\sim 5 \times 10^4$ m$^2$) 
suggests that the array should be populated
with  50-100 IACTs.

\item Fully contained events for which the shower core falls well within the geometrical
dimensions of the array,  thus giving better  angular reconstruction and much improved 
background rejection.   The performance of a typical IACT array in the energy regime below a few TeV  
is limited by the cosmic-ray background. The sensitivity of a future  
observatory could be further enhanced through improvements of its  angular  
resolution and  background rejection capabilities.  It is known that the  
 angular resolution of the present-day arrays of IACTs, which typically have  
four telescopes, is not limited by the physics of atmospheric cascades, but  
by the pixelation of their cameras and by the number of telescopes  
simultaneously observing a $\gamma$-ray event 
\cite{VF2005,Hofmann2005,FV2007}.

\item  Low energy threshold compared to existing small arrays, since contained events
           provide  sampling  of the inner light pool  where the
           Cherenkov light density is highest.  
Lower energy thresholds  (below 100~GeV) generally require larger  aperture 
($>15$ m) telescopes; however,  a  $\sim 1$ km$^2$ IACT has an intrinsic advantage
to  lower the energy  threshold due to the detection of fully contained events.

\item A wider field of view and the ability to operate the array as a survey instrument.
         
 \end{itemize}

 In order to maximize the scientific capabilities of
a $\sim 1$ km$^2$ array with respect to angular resolution, background suppression, 
energy threshold and field of view, it is necessary to study a range 
of options including the design of the individual telescopes and the array footprint.
Furthermore, it is necessary to determine the most cost effective/appropriate technology 
available. The  reliability of the individual telescopes is also a key consideration to 
minimize operating costs.

The history of the development of instrumentation for ground-based 
$\gamma$-ray astronomy has shown that a significant investment into the design  
and construction of new instruments ($\sim 10$ times the cost of previously  
existing ACTs) has yielded significant increases in sensitivity. For example,  
the construction of high resolution cameras in the 1980s assembled from  
hundreds of individual PMTs and fast electronics made the ``imaging''  
technique possible. This advancement improved the sensitivity of the observatories by  
a factor of 10 through the striking increase of angular resolution and  
cosmic-ray background rejection, and ultimately led to a detection of the  
first TeV source \cite{1989ApJ...342..379W}. Another factor of ten investment into the development of  
small arrays of mid-sized IACTs ($12$~m) demonstrated the benefits of  
``stereoscopic'' imaging and made possible the H.E.S.S. and VERITAS  
observatories. The sensitivity of these instruments improved by a factor of  
10 due to the increase of angular resolution and CR background  
discrimination, despite their only relatively modest increase in the  
$\gamma$-ray collecting area compared to the previous-generation Whipple  
$10$~m telescope. 

The next logical step in the evolution of the IACT technique is the $\sim 1$ km$^2$ array concept. 
Technological developments such as novel multi-pixel high-quantum-efficiency 
photo-detectors (MAPMTs, SiPMs, APDs, CMOS sensors, etc.) or PMTs with 
significantly improved QE, new telescope optical design(s), 
and modular low-cost electronics based on ASICs (Application-Specific Integrated Circuits)  
and intelligent trigger systems based on FPGAs (Field Programmable Gate Arrays)
hold the promise to (i) significantly reduce the price per telescope, and 
(ii) considerably improve the reliability and versatility  of IACTs. 

The improvement in sensitivity with a  $\sim 1$ km$^2$ array is in part achieved
by increasing the number of telescopes. Simple scaling suggests that a 
factor of $10^1$ improvement in  sensitivity requires a factor of $10^2$ increase
 in the number of telescopes and observatory cost.   However, this is not the case
for the $\sim 1$ km$^2$ IACT array concept, since the $\sim 1$ km$^2$ concept inherently
 provides a  better event reconstruction so that the  sensitivity improves far beyond 
 simple scaling arguments.
For the current generation of small arrays, the shower core mostly falls outside the 
physical array dimensions. 
 A $\sim 1$ km$^2$ array could, for the first time, fully constrain the air shower 
based on  many view points from the ground. This leads to several  substantial 
improvements  and can be understood by considering the
Cherenkov light density distribution at the ground. 
 
The Cherenkov light pool from an atmospheric cascade consists of three  
distinct regions: an inner region ($r<120$~m) in which the photon density is  
roughly constant, an intermediate region where density of the Cherenkov  
photons declines as a power law ($120$~m $<r<$ $300$ m) and an outer region  
where the density declines exponentially. 
A small array (VERITAS, HESS) samples the  majority of cascades in the intermediate and outer 
regions of the light pool.  A $\sim 1$ km$^2$ array samples for its mostly contained events,
 the inner, intermediate and outer region of the light pool and allows
 a much larger number of telescopes to participate in the event reconstruction with several
important consequences:

\begin{itemize}

\item First of all, at the trigger level this results in a lower energy threshold
since there are always telescopes that 
fall into the inner region where the light density is highest. For example, the 
$12$~m reflectors of the VERITAS array sample a majority of $100$ GeV 
$\gamma$ rays at distances of $\sim 160$~m
 and collect $\sim 105$ PEs per event. The same median number of photons  
would be collected by $9.3$ m reflectors, if the atmospheric cascades were  
sampled within a  distance of $~\sim 120$ m.  A $\sim 1$ km$^2$ array
 of IACTs with fully contained events could operate  effectively at energies below 
100~GeV despite having a telescope aperture  smaller than that 
of  VERITAS~\cite{VF2005,JKBF2005}.
Reducing the telescope size translates into  a reduction 
of cost per telescope and total cost for a future observatory.

\item The second factor which significantly affects the sensitivity and cost of  
future IACT arrays is the angular resolution for $\gamma$-rays. Due to the  
small footprint of the VERITAS and H.E.S.S. observatories, the majority  
of events above $\sim 100$~GeV are sampled outside the boundaries of the  
array, limiting the accuracy to which the core of atmospheric cascade can be  
triangulated.   Even higher resolution pixels will not help to improve the
angular resolution below $\sim 9$ arc-minutes ~\cite{Bugaev:07} for small
arrays.  However,  contained events in a  $\sim 1$ km$^2$ array
 of IACTs provide a nearly ideal  reconstruction based on simultaneous observations of the 
shower from all directions while sampling multiple core distances. 
Simulations of idealized  (infinite) large arrays of IACTs equipped with cameras composed  
from pixels of different angular sizes suggest that the angular resolution  
of the reconstructed arrival direction of $\gamma$-rays improves with finer  
pixelation up to the point at which the typical angular scale,  
determined by the transverse size of the shower core is  
reached~\cite{FV2007}.  Figure~\ref{fig:sensang} shows the angular  
resolution that can be achieved (few minutes of arc) with an ideal ``infinite'' array of IACTs  
when instrumental effects are neglected \cite{Hofmann2005}.

\item The third factor improving the sensitivity of  $\sim 1$ km$^2$ arrays of IACTs  
comes through enhanced background discrimination. For atmospheric cascades  
contained within the array footprint, it is possible to determine  
both the depth of the shower maximum and the cascade  
energy relatively accurately, thereby enabling better separation of hadronic and electromagnetic  
cascades. Multiple viewpoints from the ground at different core distances 
also allow the detection of fluctuations in light density and further improve background rejection.
Additional  improvements extending to energies below 200~GeV may be possible by 
picking up muons from hadronic cascades, a technique that is used in air shower 
arrays.  A ``muon veto'' signal  present in the images  obtained of a large
array could improve the technique even further. Another method to reject cosmic-ray background
at the lowest energies and low light levels \cite{FK:1995} is based  on the parallactic displacement
of images.
The images viewed from multiple viewpoints at the ground show significant fluctuations in
lateral displacements for hadronic showers and simulations indicate appreciable $\gamma$/hadron 
separation capabilities in a regime where faint Cherenkov light images can no longer be resolved
for the calculation of standard image parameters.  This technique could
become effective close to the trigger threshold of large arrays.

\end{itemize}

In summary, the concept of ``large IACT arrays'' provides strongly  
improved sensitivity at mid-energies, $\sim 1$ TeV, not only due to  
increased collecting area, but also due to enhanced angular  
resolution and CR background rejection. It also presents a 
cost-effective solution for increasing the collecting area of the  
observatory at lower energies.

For energies above $>10$~TeV, the collecting area of the $\sim 
1$ km$^2$ IACT array will be approximately two times larger than its geometrical area 
due to events impacting beyond the perimeter of  the array. 
It must be noted that in this energy regime the observatory is no  
longer background limited and therefore its sensitivity scales inversely  
proportional to the collecting area and exposure. 

Clearly, versatility is another virtue of a ``large IACT array''.  If the  
astrophysics goal is to only measure the high-energy part of the spectrum  
($>10$~TeV) of a given source, e.g. the Crab Nebulae or Galactic Center, only  
$1/10^{\mathrm{th}}$ of the observatory telescopes, spaced on the grid of  
$\sim 300$~m, would be required to participate in the study to gain a  
required sensitivity, while at the same time other observation programs 
could be conducted.   The  flexibility  of a large array also allows
operation in a sky survey mode to detect transient galactic or  
extragalactic sources~\cite{VF2005}. In this mode of operation a large field  
of view would be synthesized by partially overlapping the fields of view of  
individual telescopes. Survey observations, in which collecting  
area has been traded for wide solid-angle coverage, could then be followed up  
 by more sensitive ``narrow-field'' of view for detailed source studies.

Although the design considerations outlined above are relevant for any 
``large IACT array'',  realistic implementations of this concept could vary.
An alternative approach to the array, consisting of identical telescopes, is being developed,  
based on an extrapolation from small arrays, H.E.S.S. and  
VERITAS, and is known as the hybrid array concept.
In this approach the limitation  of the cost of the future observatory is addressed 
through a design with multiple types of IACTs, each addressing a different energy range.
For  example, a central core composed of a few very large aperture  
telescopes ($\sim 20$~m) equipped with fine pixel cameras (or very high  
spatial density mid-size reflectors~\cite{JKBF2005} ), provides for the low  
energy response of the array. A significantly larger, $\sim 1$~km$^2$, ring  
area around the array core is populated with VERITAS class telescopes  
($>12$~m) to ensure improved collecting area and performance at mid-energies,  
$\sim 1$ TeV. Finally, a third ring surrounds the 1~km$^2$ array with a very  
spread-out array of inexpensive, small ($2$~m aperture), wide-field IACTs  
outfitted with coarsely pixelated cameras ($0.25^{\circ}$), which would cover  
areas up to $10$~km$^2$. On the order of $100$ telescopes with $300$~m  
spacing might be required to gain the desired response at the highest  
energies ($> 10$ TeV)~\cite{stamatescu07}. 
 
The hybrid array concept with a central region of several large  
aperture telescopes is motivated  by  significant changes in  
the distribution of Cherenkov photons at energies considerably smaller than  
$\sim 100$ GeV.  At very low energies, $\sim  
10$ GeV, the Cherenkov light is distributed over a relatively large area, but  
with lower overall density. Therefore, large aperture telescopes arranged in  
an array with significant separation between them may provide a cost  
effective solution to improve the low energy response.

Independently from exact implementation of the IACT array layout, the  
sensitivity of future ground-based observatories could be improved through the  
increase of both camera pixelation  and the number of  
telescopes. The low energy sensitivity will also be  
affected by the telescope aperture. Therefore, a trade-off optimization of  
these factors should also be performed under a constraint of constant cost of  
the observatory. For example, if the camera dominates the overall cost of the  
IACT significantly, then a reduction of camera pixelation and increase of the  
number of telescopes is suggested for optimizing cost.
 If the telescope optical and positioning systems dominate the cost,  
then reducing the number of telescopes and improving their angular resolution  
is preferential for achieving the highest sensitivity. The cost per pixel and 
of the indivisual  telescopes of a given apearture are the most critical 
parameters required for future observatory design decisions.
 
Through the design and construction of H.E.S.S., VERITAS, and MAGIC, 
considerable experience has been gained in  
understanding the cost and technical challenges of constructing prime  
focus, Davies-Cotton (DC) and parabolic reflectors and assembling cameras  
from hundreds of individual PMTs. 
The relatively inexpensive, DC telescope design has been used in ground-based $\gamma$-ray  
astronomy for almost fifty years successfully and provides an excellent baseline option  
for a future observatory.  For example, the HESS 13~m aperture telescopes have an optical
pointspread function of better than 0.05 deg. FWHM  over a 4 degree field of view
and pixel size of 0.15~deg., demonstrating that this telescope design could in principle
accommodate a few arc minute camera resolution.
  To reach significantly better angular resolution in conjunction with wider field of view 
systems,  alternative designs are being considered.

An alternative telescope design that could be  
used in future IACT array is based on the Schwarzschild-Couder (SC) optical  
system (see Fig. \ref{fig:vass_fig2.ps})~\cite{Vass:07}, which consists of  
two mirrors configured to correct spherical and coma aberrations, and  
minimize astigmatism. For a given light-collecting area, the SC optical system  
has considerably shorter focal length than the DC optical system, and is  
compatible with small-sized, integrated photo-sensors, such as Multi Anode  
PMTs (MAPMTs) and possibly Silicon PMs (SiPMs). Although the SC telescope  
optical system, based on aspheric mirrors, is more expensive than that of a  
DC design of similar aperture and angular resolution, it offers a  
reduction in the costs of focal plane instrumentation using pixels that are physically
substantially smaller.
In addition, the SC telescope offers a wide, unvignetted, 6 degree field-of-view, 
unprecedented  for ACTs, which can be further extended up to 12 degrees, if necessary, when  
a modest degradation of imaging and loss of light-collecting area can be  
tolerated. Unlike a DC telescope, the two-mirror aplanatic SC design does not  
introduce wavefront distortions, allowing the use of fast $>$~GHz electronics to  
exploit the very short intrinsic time scale of Cherenkov light pulses ($<$3  
nsec). 
The Schwarzschild telescope design was proposed in 
1905~\cite{Schwarzschild1905}, but the construction of an SC telescope only  
became technologically possible recently due to fundamental advances in the  
process of fabricating aspheric mirrors utilizing replication processes such  
as glass slumping, electroforming, etc.  It is evident that the SC design
requires novel  technologies and  is scientifically attractive.  Prototyping
and a demonstration of its performance and cost are required to fully explore its
potential and scientific capabilities.

To summarize, ``large'' IACT array concept provides the means to achieve  
the required factor of 10 sensitivity improvement over existing instruments.
 Significant simulations and design studies are required to make an informed 
decision on the exact array implementation, such as deciding between uniform 
or graded arrays. Two  
telescope designs, DC \& SC, offer a possibility for the largest collecting  
area, largest aperture, and highest angular resolution IACT array options.  
Studies of the tradeoff of performance costs and robustness of operation are  
necessary for design conclusions.

\begin{figure*}[t] 
 
\begin{center} 
 
\includegraphics[angle=0,width=6.1cm]{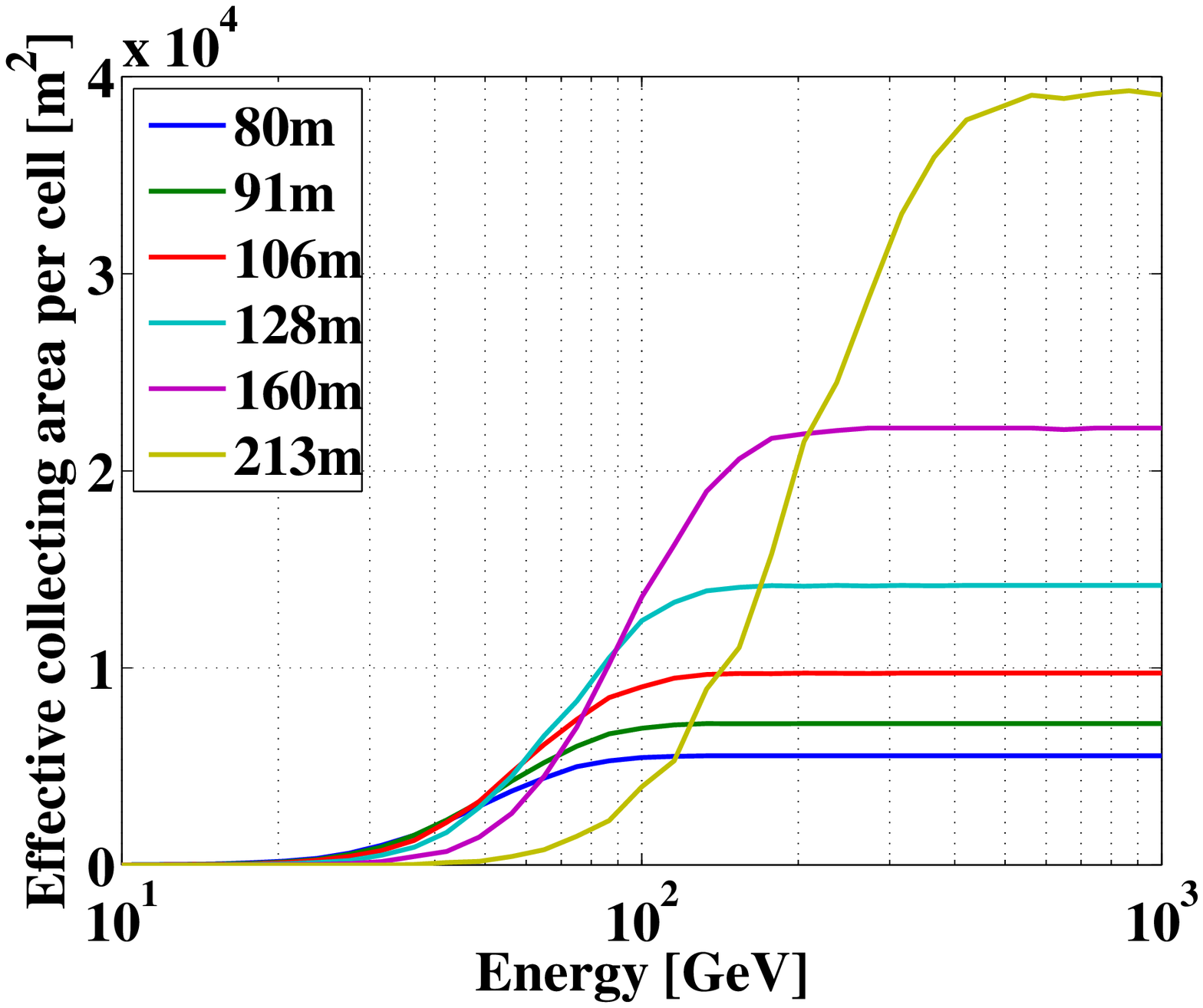} 
\includegraphics[angle=0,width=8.8cm]{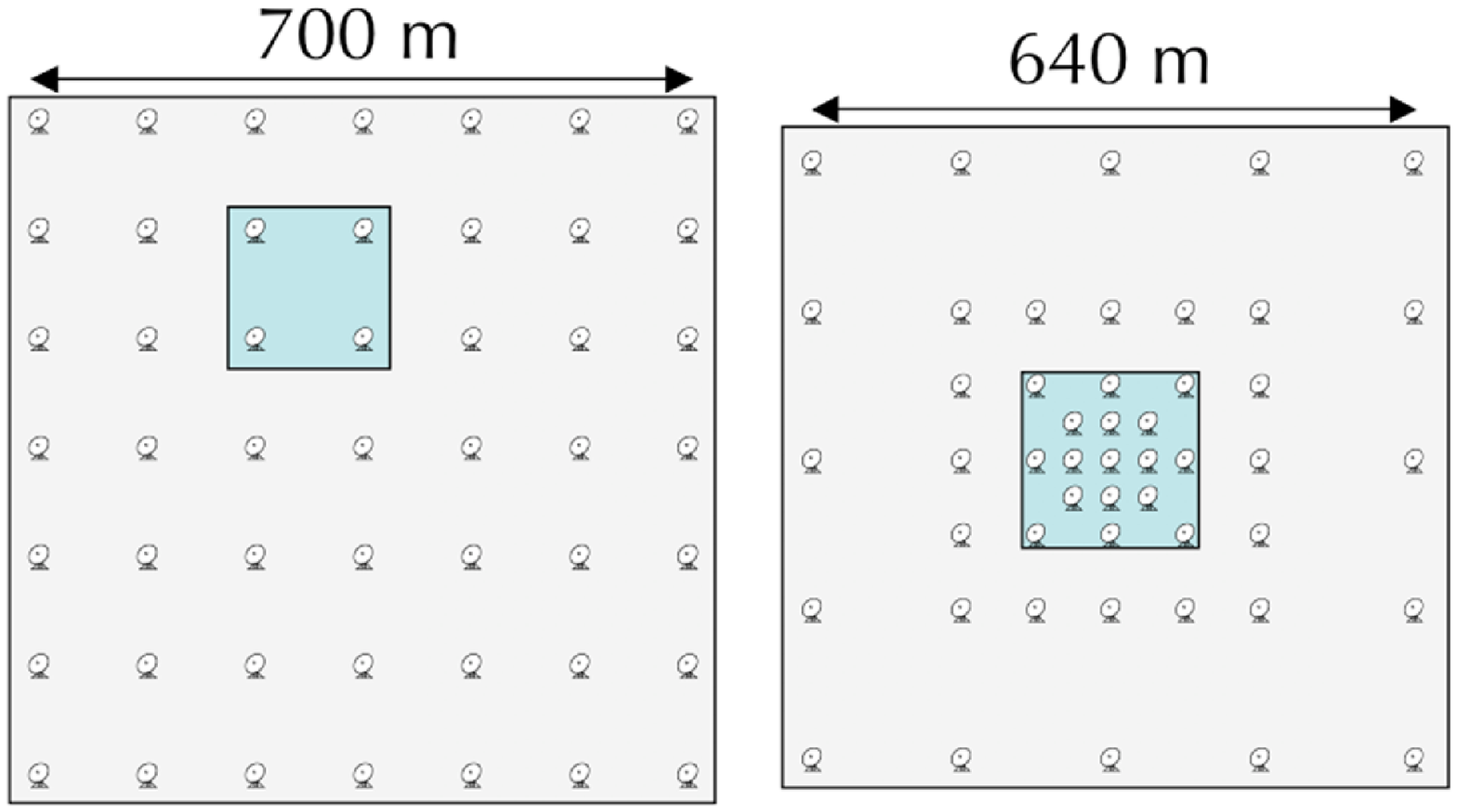} 
 
\caption{ \textit{Left:} Effective area vs. energy for a single cell for different telescope spacings; for a very large array with a fixed number of telescopes, the total effective area will be proportional to this number. \textit{Center,Right:} Two possible array configurations showing a uniform array and one where the central cluster of telescopes is more densely packed to achieve a balance between the desires for low threshold and large effective are at higher energies.}
 
\label{fig:array1}

\end{center} 
 
\end{figure*}

\begin{figure*}[t] 
 
\begin{center} 
 
 
\includegraphics[angle=0,width=4.0in]{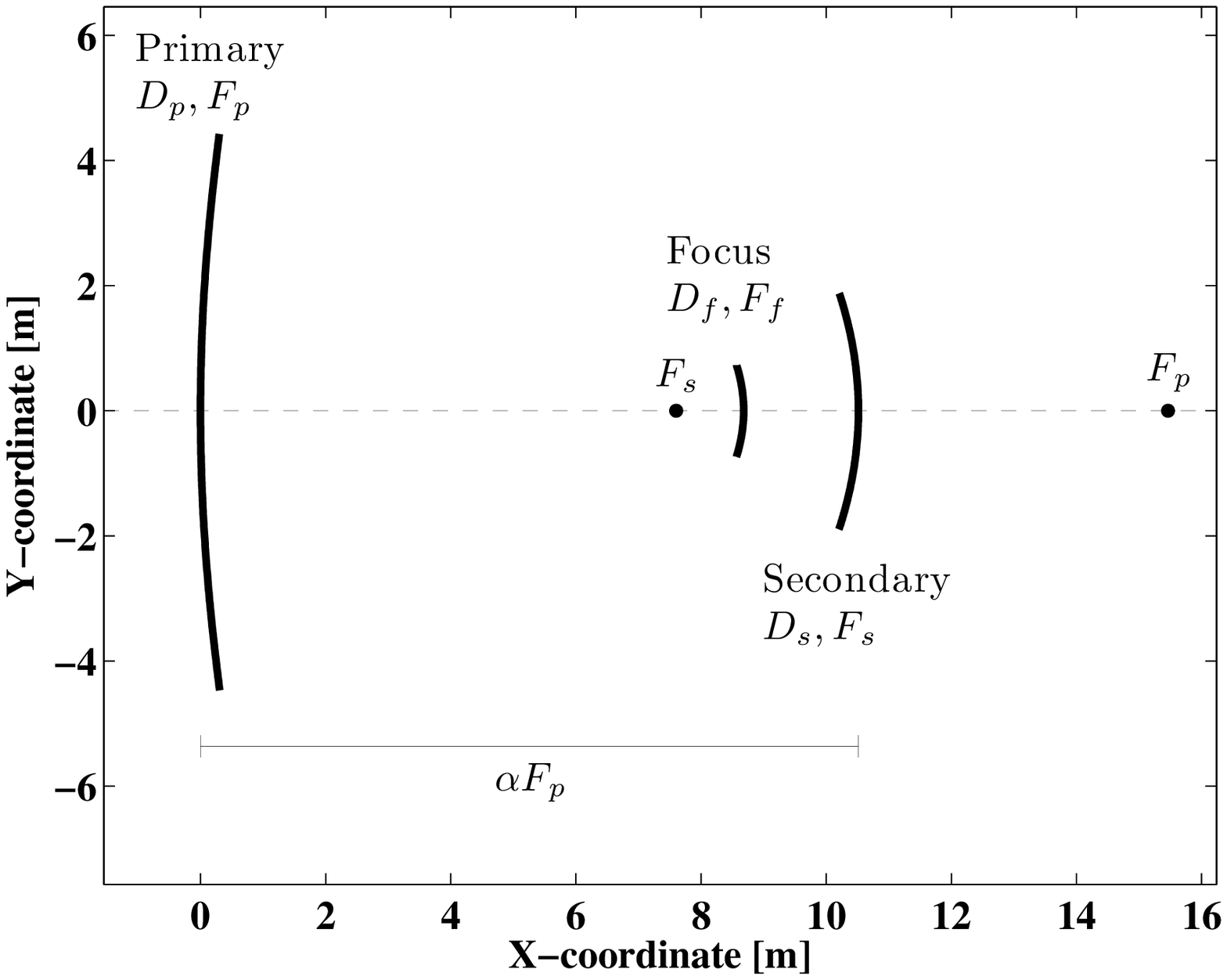} 
 
\caption{\label{fig:optics} A future Cherenkov telescope array may use  
conventional Davies-Cotton or parabolic optical reflectors 
similar to the ones used by VERITAS, MAGIC, and H.E.S.S., or may use novel  
Schwarzschild-Couder optical designs that  
combine wide field of views 
with excellent point spread functions and a reduction of the plate-scale, and  
thus of the camera size, weight, and costs. 
The image shows the cross-section of an exemplary Schwarzschild-Couder design  
(from \cite{Vass:07}).} 
\label{fig:vass_fig2.ps} 
\end{center} 

\end{figure*}

\subsection[Future EAS observatory]{Future EAS Observatory} 

The success of EAS observatories in gamma-ray astronomy is relatively recent, with the
first detection of new sources within the last couple of years \cite{Abdo:07}, as compared to the
over 20 year history of successes with IACTs.  However, EAS observatories have
unique and complementary capabilities to the IACTs. 
The strengths of the technique lie in the ability to perform unbiased all-sky surveys (not
simply of limited regions such as the Galactic plane), to measure spectra
up to the highest energies, to detect extended sources
and very extended regions of diffuse emission such as the Galactic plane, and
to monitor the sky for the brightest transient emission
from active galaxies and gamma-ray bursts and search for
unknown transient phenomena.

The instantaneous field of view of an EAS detector is $\approx$2 sr and is
limited by the increasing depth of the atmosphere that must be traversed by
the extensive air shower at larger zenith angles. 
However, for higher energy gamma rays, the showers are closer to
shower maximum and have more particles; thus the resolution improves. 
As the Earth
rotates, all sources that pass within $\approx$45 degrees of the detector's zenith
are observed for up to 6 hours.  For a source with a Crab-like spectrum, the flux sensitivity
of an EAS detector varies by less than 30\% for all sources located within $\approx$2$\pi$ sr.

The angular resolution, energy resolution, and $\gamma$-hadron separation  
capabilities of EAS technique are limited by the fact that the detectors 
sample the particles in the tail of the shower development well past the shower
maximum. The angular resolution improves  
at higher energies ($>$ 10 TeV), and the best single-photon angular  
resolution achieved to date is 0.35$^{\circ}$ which was achieved with the highest energy observations of Milagro.
Placing an extensive shower detector at a higher elevation will allow the particles
to be detected closer to the shower maximum.  For example,  an observatory at 4100m
above sea level detects 5-6 times as many particles for the same energy primary
as an observatory at 2650m (the elevation of Milagro).

Also, increasing the size of a detector will increase the collection area and thus the 
sensitivity. As both signal and background are increased, the relative sensitivity would
 scale proportional to Area$^{0.5}$ if there were no other improvements. However, the effectiveness 
of the gamma-hadron cuts improves drastically with detector size, because the lateral 
shower distribution is more thoroughly sampled. The background hadron induced showers 
can be efficiently rejected through the identification of muons, hadrons and secondary 
electromagnetic cores. But the large transverse momentum of hadronic interactions spreads 
the shower secondaries over a much larger area on the ground than the gamma-ray initiated 
showers.  Detailed simulations using Corsika to simulate the air showers and GEANT4 to simulate
a water Cherenkov observatory show that most background hadronic showers can be rejected by 
identifying large energy deposits separated from the shower core\cite{Smith_GLAST:07}. 
Simulations of larger versions of such a detector demonstrate that sensitivity scales as 
Area$^{0.8}$ at least up to 300m x 300m.

The high-energy sensitivity of all gamma-ray detectors is limited by the total exposure
because the flux of gamma rays decreases with energy.  An EAS detector has a very large
exposure from observing every source every day.  For example, a detector of area 2 $\times$ 10$^4$m$^2$ 
after 5 years will have over 1 km$^2 \times$ 100 hours of exposure. And as the energy increases, EAS
observatories become background free because the lateral distribution of muons, hadrons and secondary
cores in hadronic showers is better sampled.

The low energy response of EAS detectors is very different from IACTs, again because only the
tail of the longitudinal distribution of the shower is observed.  Past shower maximum, the 
number of particles in the shower decreases with each radiation length.  However, the 
probability of a primary penetrating several radiation lengths prior to first interaction
in the atmosphere decreases exponentially with radiation length.  These two facts, as well as
the number of particles at shower maximum is proportional to the primary energy,
imply the effective area increases with energy E as E$^{2.6}$ until a threshold energy 
where the shower can be detected if the primary interacts within the first radiation
length in the atmosphere. Therefore, EAS detectors can have an effective area up to 100 m$^2$ at 
the low energies of $\sim$ 100 GeV.  This area is considerably larger than Fermi's of $\sim$ 1
m$^2$, and is sufficient to observe bright, extragalactic sources such as active
galactic nuclei and possibly gamma-ray bursts.  The wide field of view of EAS observatories
is required to obtain long term monitoring of these transient sources and EAS observatories
search their data in real time for these transient events to send notifications within a few
seconds to IACTs and observers at other wavelengths.

The HAWC (High Altitude Water Cherenkov) observatory is a next logical step in the development
of EAS observatories\cite{dingus:07}.  It will be located in Mexico at Sierra Negra at an altitude of 4100 m and
will have 10-15 times the sensitivity of Milagro. The (HAWC) observatory will re-use the 
existing photomultiplier tubes from Milagro in an approximately square array of 900 large water tanks.  The tanks
will be made of plastic similar to the Auger tanks, but will be larger, with a diameter of 5 m and
4.3 m tall. An 8" diameter PMT would be placed at the bottom of each tank and look up
into the water volume under $\approx$4 m of water.
The array would enclose 22,500 m$^2$ with $\approx$75\% active area.
Thus, unlike Milagro, the same layer of
PMTs would be used to both reconstruct the direction of the primary gamma ray
and to discriminate against the cosmic-ray background.   
The optical isolation of each PMT in a separate tank allows a single layer to accomplish
both objectives. A single tank has been tested in conjunction with Milagro and its 
performance agrees with Monte Carlo simulation predictions.  The optical isolation also
improves the background discrimination (especially at the trigger level), and the 
angular and energy resolution of the detector.

\begin{figure*}[ht] 
\begin{center}
\includegraphics[height=5.2in]{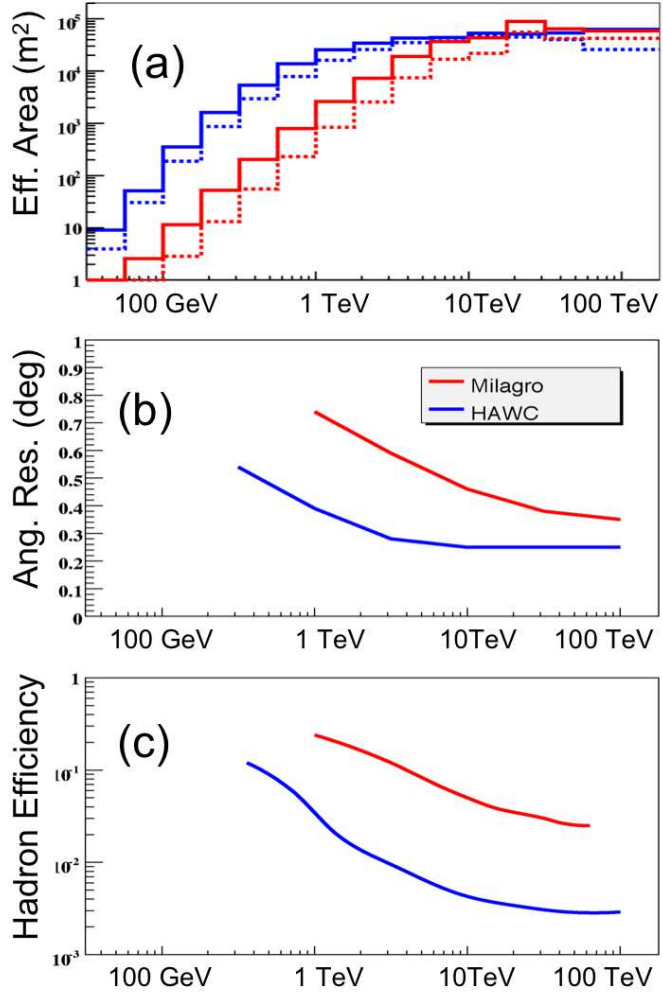}  
\caption{\label{fig:hawc}: The sensitivity of HAWC and Milagro versus primary gamma-ray energy. 
Panel (a) shows the effective area, (b) the angular resolution, and (c) the efficiency with the hadronic background showers are rejected when half of the gamma-ray events are accepted. }
\end{center}  
\end{figure*}

The performance of HAWC is shown in Figure \ref{fig:hawc} and is compared to Milagro.  These detailed
calculations use the same Monte Carlo simulations that accurateley predict the performance of Milagro.
The top panel shows the large increase in the effective area at lower energies as expected from the 
increase in altitude from 2600m to 4100m.  At higher energies the geometric area of HAWC is similar
to the geometric area of Milagro with its outrigger tanks.  However, the improved sampling of the
showers over this area with the continuous array of HAWC tanks results in improved angular resolution
and a major increase in background rejection efficiency.  Therefore, the combined sensitivity improvement 
for a Crab-like source is a factor of 10-15 times better than Milagro.  This implies that the Crab 
can be detected in one day as compared to three months with Milagro.   

The water Cherenkov EAS detector  can be extrapolated to enclose even larger areas and
the sensitivity of such a detector is relatively straight forward to calculate.
Earlier work in this area discussed an array enclosing 100,000 m$^2$, with
two layers of PMTs \cite{Sinnis2004, Sinnis2005}.  Recent work
indicates that a single deep layer (as in the HAWC design)
will perform as well as the previous two-layer design. 
For example, a detector with an active detection
area 100,000 m$^2$ (HAWC100), located at 5200 m above sea level, would have an effective
area at 100 GeV of $\sim$10,000 m$^2$ for showers from zenith.
The low-energy response allows for the detection of gamma-ray bursts at larger
redshifts than current instruments ($z\sim$1 for HAWC compared to $z\sim$0.3
for Milagro if, at the source, the TeV fluence is equal to the keV fluence).
While current instruments, such as Milagro, indicate that the
typical TeV fluence from a GRB is less than the keV fluence, instruments such as
HAWC100 and HAWC would be sensitive to a TeV fluence 2-3 orders of magnitude
smaller than the keV fluence of the brightest gamma-ray bursts.

\subsection[Technology roadmap]{Technology Roadmap}
\label{sec:RoadMap}
The recent successes of TeV $\gamma$-ray astronomy both in terms of  
scientific  
accomplishments and in terms of instrument performance have generated 
considerable interest in next-generation instruments. Part of the excitement  
originates from the fact that an order of magnitude sensitivity improvement  
seems to be in reach and at acceptable costs for making use of existing technologies.  
New technologies could result in even better sensitivity improvements.
A roadmap for IACT instruments over 
the next 3 years should focus on design studies to understand the trade-
offs between performance, costs, reliability of operation of IACT arrays, 
and on carrying out prototyping and the required research and development. 
It is anticipated that, at the end of this R\&D phase, a full proposal for construction of an observatory would be submitted.  
A next generation instrument could be built on a time scale of $\sim$5 years  
to then be operated for between 5 years (experiment-style operation) and  
several decades (observatory-style operation). 
For IACT instruments, the following R\&D should be performed: 
\begin{itemize}

\item Monte Carlo simulations of performance of large IACT arrays to optimize array configuration parameters such as array type (hybrid or homogeneous), array layout, aperture(s) of the telescope(s), and pixilation of the cameras, with a fixed cost constraint.  Effects of these parameters on energy threshold, angular resolution, and sensitivity of the observatory should be fully understood, together with associated cost implications.

\item The conservative Davies-Cotton telescope design with $f - \frac{F}{D} \sim 1$ should be considered as 
a baseline option for the future observatory.  However, limitations of this design and benefits and cost impact of alternative options should be investigated.  These alternatives include large focal length Davies-Cotton or parabolic prime-focus reflectors with $f\sim 2$ and aplanatic two-mirror optical systems, such as Schwarzschild-Couder and Ritchey-Chr\'{e}tien telescopes.  The latter designs have the potential to combine significantly improved off-axis point spread functions, large field-of-views, and isochronicity with reduced plate scales and consequently reduced costs of focal plane instrumentation.  Prototyping of elements of the optical system of SC or RC telescopes is required to assess cost, reliability and performance improvement.  Mechanical engineering feasibility studies of large focal length prime focus telescopes and two-mirror telescopes should be conducted.

\item The development and evaluation of different camera options should be  
continued. Of particular interest  
are alternative photo-detectors (photomultiplier tubes with ultra high  
quantum efficiency,  
multi-anode photomultipliers, multi channel plates, Si photomultipliers,  
Geiger mode Si detectors,  
and hybrid photodetectors with semiconductor photocathodes such as GaAsP or  
InGaN) and a modular design  
of the camera which reduces the assembly and maintenance costs. Compatibility of these options with different telescope designs and reliability of operation and cost impact should be evaluated.

\item The development of ASIC-based front-end-electronics should be continued  
to further minimize the power  
and price of the readout per pixel. 

\item A next-generation experiment should offer the flexibility to operate in  
different configurations, so that specific 
telescope combinations can be used to achieve certain science objectives.  
Such a system requires the development of 
a flexible trigger system.  Furthermore, the R\&D should explore the possibility of combining the trigger signals of  
closely spaced telescopes to synthesize a single telescope of larger  
aperture. A smart trigger could be used  
to reduce various backgrounds based on parallactic displacements of Cherenkov light images \cite{FK:1995}. 

\item The telescope design has to be optimized to allow for mass production  
and to minimize the maintenance costs. 

\item The telescopes should largely run in robotic operation mode to enable a  
small crew to operate the entire system.  The reliability of operation of large IACT arrays should be specifically researched, including tests of instrumentation failure rates and weathering to evaluate required maintenance costs.

\end{itemize}

A roadmap for EAS array over the next 5 years (HAWC) is well defined by the benefits of moving the experiment to high altitudes and   
enlarging the detection area.  The cost of this path is  $<$ \$10M USD.  A site in Mexico has been identified and is a few km from the Large Millimeter Telescope; it is a 2 hour drive from the international airport in Puebla, and has existing infrastructure of roads, electricity, and internet.  The HAWC project will be a joint US and Mexican collaboration with scientists from Milagro, Auger, and other astronomical and high-energy physics projects.

The R\&D for IACT could be finalized on a time scale of  
between 3 (IACTs).  
The R\&D should go hand in hand with the establishment of a suitable  
experimental site and the build-up of basic infrastructure. 
Ideally, the site should offer an easily accessible area exceeding 1 km$^2$.  
For an IACT array, an altitude between 2 km and 3.5 km will give the best  
tradeoff between low energy thresholds,  
excellent high-energy sensitivity, and ease of construction and operation.

The U.S.\ teams have pioneered the field of ground based $\gamma$-ray  
astronomy during the last 50 years. The U.S. community has formed the  
``AGIS'' collaboration  
(Advanced Gamma ray Imaging System) to  optimize the design of a future  
$\gamma$-ray detector.  
A similar effort is currently under consideration in Europe by the CTA  
(Cherenkov Telescope Array) group, and the 
Japanese/Australian groups building CANGAROO are also exploring avenues for  
future progress. 
Given the scope of a next-generation experiment, the close collaboration of  
the US teams with the European and  
Japanese/Australian groups should be continued and intensified. If funded  
appropriately, the US teams are in  
an excellent position to lead the field to new heights.